\documentclass[aps,pra,twocolumn,showpacs,superscriptaddress,groupedaddress]{revtex4-1}
\usepackage{graphicx}    
\usepackage{dcolumn}    
\usepackage{bm}             
\usepackage{amssymb}   
\usepackage{amsmath}    
\usepackage{mathrsfs}    
\usepackage{commath}   
\usepackage{subfigure}    
\usepackage{braket}
\usepackage{natbib}
\usepackage{float}
\begin{document}

\title{Decoherence of rotational degrees of freedom}
\author{Changchun Zhong}
\email{zchangch@purdue.edu} \affiliation{Department of Physics and Astronomy, Purdue University, West Lafayette IN, 47907 USA}\affiliation{ Purdue Quantum Center, Purdue University, West Lafayette, IN, 47907, USA}
\author{F. Robicheaux}
\email{robichf@purdue.edu} \affiliation{Department of Physics and Astronomy, Purdue University, West Lafayette IN, 47907 USA}\affiliation{ Purdue Quantum Center, Purdue University, West Lafayette, IN, 47907, USA}

\date{\today}

\begin{abstract}

The mechanism of decoherence for a quantum system with rotational degrees of freedom is studied. From a simple model of elastic scattering, we show that the non-diagonal density matrix elements of the system exponentially decay. The decay rate depends on the difference of scattering amplitudes for different rotational configurations, leading to the gradual loss of quantum coherence between the pointer states in the system orientational space. For a dielectric ellipsoid immersed in a photon-gas environment (assuming no absorption), the decay rate is found to be proportional to the seventh power of the temperature. For an ellipsoidal object interacting with mass particles, the decay rate is proportional to the 5/2 power of the temperature. Both are different from the case of translational decoherence induced by the same environment scattering. For photon scattering, the coherence time in the rotational degrees of freedom is shown to be much shorter than that in the translational degrees of freedom.

\end{abstract}
\maketitle

\section{introduction}

Decoherence refers to the mechanism through which the classical world effectively emerges from quantum systems. It is a process of a quantum system losing its quantum coherence between pointer states. Decoherence is not only of great importance to the foundations of quantum physics \cite{EJ,MS,WHZ1,WHZ2}, but also of vital interests for the realization of quantum applications, such as quantum computers \cite{NC}, and other coherent manipulations \cite{CRPS} etc. The key idea of decoherence is very simple and clear, that a quantum system in reality is essentially open because it could never be completely isolated from its environment. Thus, this open quantum system is not expected to follow the Schr$\ddot{\text{o}}$dinger equation, instead, it evolves non-unitarily according to a master equation \cite{EJ,MS}, which allows a natural description of decoherence.
 
The past several decades have seen enormous progress in the area of decoherence. First, decoherence rules out almost any possibility of everyday-macroscopic-scale coherence. When the interaction with the environment is taken into account, a macroscopic system generally suffers from the loss of quantum coherence in an extremely short time. Interestingly, the loss of quantum coherence selectively happens between a certain set of basis, which is known as the pointer basis \cite{WHZ1,WHZ2,WHZ3}. The corresponding density matrix expressed in the pointer basis is quickly reduced to a classical mixed distribution. This process (known as super-selection) is guaranteed by the environment continuously monitoring certain observables of the system, which is essential for the understanding of quantum to classical transitions \cite{WHZ3}. Second, for different systems, much effort has been devoted to specifying the decoherence mechanisms, as well as to finding a way to minimize the decoherence rate, which is important for accurate quantum control. With fast technological developments in all areas of physics, many new experiments become possible for the creation and control of quantum superposition states, from the microscopic to the mesoscopic scale, even living organisms \cite{GSES,LCSD,YYJ,YCJ,ROJM,TZ}. Thus, understanding the role decoherence plays is not only fundamentally profound, but also of practical urgency. 

As quantum phenomena become observable at larger and larger scales \cite{GSES, LCSD,YYJ}, a system's rotational or vibrational degree of freedom becomes significant. However, in the past decades, most of the attention has been concentrated on the translational degrees of freedom of a system \cite{MS,AS,HS,ASL}, where environmental interactions can produce exponentially localized wave packets. Thus, natural questions could be asked whether a similar behavior exists for a system's internal degrees of freedom, and what is the possible relation of decoherence in different degrees of freedom. The answer to these questions should provide a guidance in the growing number of experiments probing and controlling the internal degrees of freedom of a system. 

In Ref. \cite{TF}, Fischer considered the interaction of a rigid body with an environment through the coupling of a dipole moment and a fluctuating field. Based on the method of quantum stochastic differential equation (SDE), Fischer derived a master equation
\begin{equation}
\partial_t\rho\propto\int_0^\infty ds\iint d\hat{n}_1d\hat{n}_2\gamma(s)(e^{if_{\bar{s}}(\hat{\Omega})}\rho e^{-if_{\bar{s}}(\hat{\Omega})}-\rho),
\end{equation}
which describes the dynamics of an extended object interacting with random pulsed fields (the unitary part is not included). The master equation gives an exponential decay of angular coherence, with the decay rate proportional to the distance measure in orientational space. In Ref. \cite{TF}, it is worth mentioning that the master equation was also used to discuss the pointer states of the orientational decoherence, and solitonic solutions were identified as the pointer states, which is instrumental for the analysis of rotational dynamics. 

In this paper, we extend the first results of Ref. \cite{TF} to include rotational decoherence due to interaction with an environment of discrete particles. The derivation is based on a model of scattering which has been widely applied to the decoherence for translational degrees of freedom \cite{EJ,MS,HS,AS,BMH}. The single scattering event is treated in a non-perturbative way \cite{MS,AS,HS}. If the system is initially prepared in a rotational superposition state, the coherence in the density matrix is effectively decreased by scattering. For a spherically symmetric environment, the rotational decoherence rate is found to depend only on the difference of the scattering amplitudes for different rotational configurations and can be shown to only depend on the difference in the angles of orientation. To illustrate the effect, we calculate the decoherence rate for the situation with the environment being a photon gas or being mass particles. In the long wavelength limit, the rotational decoherence rate is found to have a temperature dependence different from that for translational decoherence. For photon scattering by systems of the same size, the rotational decoherence rate is shown to be much larger than that in the translational degrees of freedom.

In the sections that follow, we first introduce the derivation of rotational decoherence, present the examples of a system immersed in a photon gas or mass particles environment, compare the result to our expectations, and comment on their implications and possible guidance for future experiments.

\section{Rotational decoherence due to scattering}
We focus on the rotational degrees of freedom of an object (system $S$), which interacts with the environment (environment $E$), and they are assumed to be initially uncorrelated. The combined system ($SE$) is described by a product state
\begin{equation}
\hat{\rho}_{SE}=\hat{\rho}_{S}\otimes \hat{\rho}_{E},
\end{equation}
where $\rho_{E}$ is the environment density operator, and $\rho_{(S)}$ is the system density operator. We denote the orientation eigenstate of the system by $\ket{\Omega}=\ket{\alpha,\beta,\gamma}$ \cite{TF,TF2}, where $\alpha,\beta$, and $\gamma$ are the Euler angles. In the orientation space, the system density operator takes the form
\begin{equation}
\hat{\rho}_S=\int d\Omega \int d\Omega^\prime \rho_S(\Omega,\Omega^\prime)\ket{\Omega}\bra{\Omega^\prime},
\end{equation}
and the environment density operator is taken to be
\begin{equation}
\hat{\rho}_E=\int d^3\vec{k}\rho_E(\vec{k})\ket{\vec{k}}\bra{\vec{k}}.
\end{equation}
The environment particles are initially uncorrelated such that all the non-diagonal matrix elements are zero. In the following, we will show how the system state is affected by a single scattering event. The derivation formally follows the steps of the model of collisional decoherence \cite{EJ,EH,MS}.

\subsection{Correlation established by scattering}

Our discussion is confined to elastic scattering throughout the paper. The state $\ket{\Omega}=\ket{\alpha,\beta,\gamma}$ can be represented as a state $\ket{0,0,0}$ rotated by an operator
\begin{equation}
\ket{\Omega}=\hat{D}_S(\Omega)\ket{0,0,0},
\end{equation}
where $\hat{D}_S(\Omega)=\exp(-\frac{i}{\hbar}\hat{L}_z\alpha)\exp(-\frac{i}{\hbar}\hat{L}_y\beta)\exp(-\frac{i}{\hbar}\hat{L}_z\gamma)$ \cite{SKR}.  If we denote the incoming particle by $\ket{\chi}$, then the effect of the scattering event can be described by the scattering operator $\hat{S}$ acting on the initial state,
\begin{equation}
\ket{\Omega}\ket{\chi}\rightarrow\hat{S}\ket{\Omega}\ket{\chi}.
\end{equation}
Then we have,
\begin{equation}
\begin{split}
\hat{S}\ket{\Omega}\ket{\chi} & \rightarrow \hat{S}\hat{D}_{SE}(\Omega)\ket{0,0,0}\hat{D}^\dag_E(\Omega)\ket{\chi}\\
& \rightarrow \hat{D}_{SE}(\Omega)\hat{S}\ket{0,0,0}\hat{D}^\dag_E(\Omega)\ket{\chi},
\end{split}
\end{equation}
where $\hat{D}_E(\Omega)$ is the rotational operator acting only on the environment and $\hat{D}_{SE}(\Omega)$ is the rotational operator for the combined system. Obviously, we have $\hat{D}_{SE}(\Omega)=\hat{D}_S(\Omega)\hat{D}_E(\Omega)$. In the last line, we used the fact that the scattering operator commutes with the rotation of the combined system
\begin{equation}
[\hat{S},\hat{D}_{SE}(\Omega)]=0.
\end{equation}
In the scattering model, an important step is to include the non-recoil approximation, which states that the scattering event essentially does not disturb the system, except establishing entanglement between the system and the incoming particle \cite{MS}. In the situations we are considering, the system is much more massive than the environment particles, such as photons scattered by a mesoscopic dielectric or air molecules scattered by more massive objects. This justifies the non-recoil approximation, which gives
\begin{equation}
\label{e6}
\begin{split}
\hat{S}\ket{\Omega}\ket{\chi} &\rightarrow \hat{D}_{SE}(\Omega)\ket{0,0,0}\hat{S}\hat{D}^\dag_E(\Omega)\ket{\chi} \\
& \rightarrow \ket{\Omega}\hat{S}(\Omega)\ket{\chi}\rightarrow \ket{\Omega}\ket{\chi(\Omega)},
\end{split}
\end{equation}
where $\hat{S}(\Omega)=\hat{D}_E(\Omega)\hat{S}\hat{D}^\dag_E(\Omega)$. $\ket{\chi(\Omega)}=\hat{S}(\Omega)\ket{\chi}$ is introduced to denote the state of the outgoing particle, which now carries the orientational information of the system. The first line used the non-recoil approximation. From Eq. (\ref{e6}), we see the scattering event establishes correlations between the system and the environment. Accordingly, the initial separable density matrix of the combined system $\rho_{SE}=\rho_{S}\otimes\rho_{E}$ is transformed into the following entangled density matrix
\begin{equation}
\rho_{SE}=\int d\Omega\int d\Omega^\prime\rho_S{(\Omega,\Omega^\prime)}\ket{\Omega}\bra{\Omega^\prime}\otimes\ket{\chi{(\Omega)}}\bra{\chi{(\Omega^\prime)}}.
\end{equation}
The system is described by a reduced density matrix $\rho_{S}$, which is obtained by tracing over the environmental degree of freedom, 
\begin{equation}
tr_E(\rho_{SE})=\int d\Omega\int d\Omega^\prime\rho_E(\Omega,\Omega^\prime)\ket{\Omega}\bra{\Omega^\prime}\braket{\chi(\Omega^\prime)|\chi(\Omega)}.
\end{equation}
As a result, the density matrix element of the system after the scattering event becomes
\begin{equation}
\label{e8}
\rho_S(\Omega,\Omega^\prime,0)\rightarrow\rho_S(\Omega,\Omega^\prime,0)\braket{\chi(\Omega^\prime)|\chi(\Omega)},
\end{equation}
where $\braket{\chi(\Omega^\prime)|\chi(\Omega)}=\braket{\chi|\hat{S}^\dag(\Omega^\prime)\hat{S}(\Omega)|\chi}$. Thus, a suppression is attached to the system density matrix elements, and the value is determined by the average of the operator $\hat{S}^\dag(\Omega^\prime)\hat{S}(\Omega)$ over the state of the incoming particle. The overlap $\braket{\chi(\Omega^\prime)|\chi(\Omega)}$ is trivially one for $\Omega = \Omega^\prime$, which indicates no influence on the diagonal elements from the scattering.

\subsection{Time evolution of the system density matrix}

To derive how the system density matrix evolves in time, we first need to calculate the overlap $\braket{\chi(\Omega^\prime)|\chi(\Omega)}$ and then determine how the system density matrix is affected by successive scattering events. According to Eq. (\ref{e8}), The suppression of the system density matrix element is determined by the average of the operator $\hat{S}^\dag(\Omega^\prime)\hat{S}(\Omega)$ in terms of the incoming environment particles. To calculate this average, the state of the incoming particle needs to be specified. We first confine the environment particle in a box with periodic boundary conditions. The box volume has a finite value $V$ and the momentum eigenstate in this space is denoted by $\ket{\vec{K}}$. Then we push the box size to the limit of infinity, such that the momentum eigenstate becomes continuous and is denoted by $\ket{\vec{k}}$. Considering the normalization condition, these eigenstates have the following simple connection
\begin{equation}
\label{e9}
\ket{\vec{K}}=\sqrt{\left(\frac{(2\pi)^3}{V}\right)}\ket{\vec{k}},\frac{(2\pi)^3}{V}\sum=\int d^3\vec{k}.
\end{equation}
Thus, the state of the incoming particle is described by the density operator,
\begin{equation}
\rho_E=\frac{(2\pi)^3}{V}\sum\mu(\vec{k})\ket{\vec{K}}\bra{\vec{K}},
\end{equation}
where the summation runs over the set of momenta that satisfy the periodic boundary condition. $\mu(\vec{k})$ is the wave number distribution. We assume that the environment is spherically symmetric such that $\mu(\vec{k})$ depends only on the magnitude of $\vec{k}$. Then the average of the operator $\hat{S}^\dag(\Omega^\prime)\hat{S}(\Omega)$ can be written as
\begin{equation}
\braket{\chi|\hat{S}^\dag(\Omega^\prime)\hat{S}(\Omega)|\chi} \rightarrow \frac{(2\pi)^3}{V}\sum \mu(k)\bra{\vec{K}}\hat{S}^\dag(\Omega^\prime)\hat{S}(\Omega)\ket{\vec{K}}.
\end{equation}
To proceed, the identity $\hat{S}=\hat{I}+i\hat{T}$ is used to express the scattering operator $\hat{S}$ in terms of $\hat{T}$ operator. Recall the definition $\hat{S}(\Omega)=\hat{D_E}(\Omega)\hat{S}\hat{D}^\dag_E(\Omega)$, the above expression is written as
\begin{widetext}
\begin{equation}
\label{e13}
\begin{split}
\braket{\chi(\Omega^\prime)|\chi(\Omega)} \rightarrow 1-\frac{(2\pi)^3}{V}\int d^3\vec{k}\mu(k)\bra{\vec{k}}\hat{T}^\dag\hat{T}-D_E(\Omega^\prime)\hat{T}^\dag D^\dag_E(\Omega^\prime)D_E(\Omega)\hat{T}D^\dag_E(\Omega)\ket{\vec{k}},
\end{split}
\end{equation}
\end{widetext}
where the identities $-i\hat{T}^\dag+i\hat{T}=-\hat{T}^\dag\hat{T}$, $\int d^3\vec{k}\mu(k)=1$ and Eq. (\ref{e9}) are used. While obtaining the above expression, we also used the fact that the environment is spherically symmetric, which is equivalent to state that the environment density operator $\rho_E$ commutes with the environmental rotation $D_E(\Omega)$. For the same reason, the above expression can be written in a more symmetric form
\begin{widetext}
\begin{equation}
\begin{split}
\braket{\chi(\Omega^\prime)|\chi(\Omega)} \rightarrow& 1-\frac{(2\pi)^3}{2V}\int d^3\vec{k}\mu(k)\bra{\vec{k}}\hat{T}^\dag_\Omega\hat{T}_\Omega+\hat{T}^\dag_{\Omega^\prime} \hat{T}_{\Omega^\prime}-\hat{T}^\dag_{\Omega^\prime}\hat{T}_\Omega-\hat{T}^\dag_{\Omega^\prime} \hat{T}_\Omega\ket{\vec{k}},
\end{split}
\end{equation}
\end{widetext}
where we denote $\hat{T}_\Omega=D_E(\Omega)\hat{T}D^\dag_E(\Omega)$, which is the rotated $\hat{T}$ operator. Next, we connect the $\hat{T}$ operator with the scattering amplitude by the following familiar formula \cite{SKR}
\begin{equation}
\bra{\vec{k}}\hat{T}_\Omega\ket{\vec{k^\prime}}=-\frac{\hbar^2}{2\pi m}\delta(E-E^\prime)f_\Omega(k\hat{k},k\hat{k^\prime}),
\end{equation}
and use the identity operator $\hat{I}=\int d^3\vec{k}^\prime\ket{\vec{k}^\prime}\bra{\vec{k}^\prime}$. After several steps of algebra, we obtain
\begin{widetext}
\begin{equation}
\begin{split}
\braket{\chi(\Omega^\prime)|\chi(\Omega)} \rightarrow&1-\frac{(2\pi)^3}{2V}\int d^3\vec{k^\prime}\int d^3\vec{k}\mu(k)\frac{\hbar^4}{(2\pi m)^2}\delta^2(E-E^\prime)\bigg\{ f^\ast_\Omega(\vec{k^\prime},\vec{k}) f_\Omega(\vec{k}^\prime,\vec{k})+f^\ast_{\Omega^\prime}(\vec{k}^\prime,\vec{k})f_{\Omega^\prime}(\vec{k}^\prime,\vec{k})\\
&-f^\ast_{\Omega^\prime}(\vec{k}^\prime,\vec{k})f_{\Omega}(\vec{k}^\prime,\vec{k})-f^\ast_{\Omega^\prime}(\vec{k}^\prime,\vec{k})f_{\Omega}(\vec{k}^\prime,\vec{k}) \bigg\},
\end{split}
\end{equation}
\end{widetext}
where we are encountered with a squared delta function. Inspired by the usual approach in deriving the Fermi's Golden rule, the squared delta function can be evaluated by the following formula \cite{MS,HS,BMH},
\begin{equation}
\label{e18}
\delta^2(E^\prime-E)=\frac{t}{2\pi\hbar}\delta(E^\prime-E)=\frac{t}{2\pi\hbar}\frac{m}{\hbar^2k}\delta(k^\prime-k),
\end{equation}
where the parameter $t$ is interpreted as the time when the interaction is on during the scattering event and is assumed to be much shorter than the system's decoherence time induced by a large number of collisions \cite{MS}. Using Eq. (\ref{e18}) and integrating the magnitude of momentum $k^\prime$, we get 
\begin{widetext}
\begin{equation}
\label{e19}
\begin{split}
\braket{\chi(\Omega^\prime)|\chi(\Omega)}\rightarrow& 1-\frac{t}{2V}\int dk k^2\mu(k)\frac{\hbar k}{m}\iint d^2\hat{k}d^2\hat{k}^\prime\bigg\{ f^\ast_\Omega(k\hat{k}^\prime,k\hat{k}) f_\Omega(k\hat{k}^\prime,k\hat{k})+f^\ast_{\Omega^\prime}(k\hat{k}^\prime,k\hat{k})f_{\Omega^\prime}(k\hat{k}^\prime,k\hat{k})\\
&-f^\ast_{\Omega^\prime}(k\hat{k}^\prime,k\hat{k})f_{\Omega}(k\hat{k}^\prime,k\hat{k})-f^\ast_{\Omega^\prime}(k\hat{k}^\prime,k\hat{k})f_{\Omega}(k\hat{k}^\prime,k\hat{k}) \bigg\}.
\end{split}
\end{equation}
\end{widetext}
The above expression in the integral can be further simplified. Since the scattering amplitude satisfies $f_{\Omega}(k\hat{k}^\prime,k\hat{k})=f^\ast_{\Omega}(k\hat{k},k\hat{k}^\prime)$ and $k\hat{k},k\hat{k}^\prime$ are symmetric in swapping the integral index, the double solid angle integral of each term is real. It means $\iint d^2\hat{k}d^2\hat{k^\prime}f^\ast_{\Omega^\prime}(k\hat{k}^\prime,k\hat{k})f_{\Omega}(k\hat{k}^\prime,k\hat{k})=\iint d^2\hat{k}d^2\hat{k^\prime}f_{\Omega^\prime}(k\hat{k}^\prime,k\hat{k})f^\ast_{\Omega}(k\hat{k}^\prime,k\hat{k})$. Thus the above formula can be expressed in a more symmetric form,
\begin{widetext}
\begin{equation}
\label{e20}
\begin{split}
\braket{\chi(\Omega^\prime)|\chi(\Omega)}\rightarrow &1-\frac{t}{2V}\int dk k^2\mu(k)\frac{\hbar k}{m}\iint d^2\hat{k}d^2\hat{k^\prime} \abs{ f_\Omega(k\hat{k^\prime},k\hat{k})-f_{\Omega^\prime}(k\hat{k^\prime},k\hat{k}) }^2 . 
\end{split}
\end{equation}
\end{widetext}
Equation (\ref{e20}) gives the suppression of the system density matrix element by one single elastic scattering event. The result depends on the difference of the elastic scattering amplitudes for different orientations. When $\Omega=\Omega^\prime$, the overlap is trivially one, which indicates no suppression on the diagonal elements of the system density matrix. Next we can proceed to derive the time evolution of the system density matrix. By substituting the above result into Eq. (\ref{e8}) and taking the limit $t\rightarrow0$, the following formula is obtained
\begin{equation}
\label{e21}
\frac{\partial\rho_S({\Omega,\Omega^\prime,t})}{\partial t}=-\Lambda\ast\rho_S(\Omega,\Omega^\prime,t),
\end{equation}
where the factor
\begin{widetext}
\begin{equation}
\Lambda=\frac{1}{2V}\int dk k^2\mu(k)\frac{\hbar k}{m}\iint d^2\hat{k}d^2\hat{k^\prime} | f_\Omega(k\hat{k^\prime},k\hat{k})-f_{\Omega^\prime}(k\hat{k^\prime},k\hat{k}) |^2.
\end{equation}
\end{widetext}
The above expression shows an exponential decay in the off diagonal elements of the system density matrix. $\Lambda$ is the decay rate. Eq. (\ref{e21}) describes the decoherence effect by one environment particle scattering. An ensemble of $N$ particles will build up the decoherence effect in a way that the decoherence rate is multiplied by the number of particles $N$, thus
\begin{widetext}
\begin{equation}
\label{e22}
\Lambda=\frac{N}{2V}\int dk k^2\mu(k)\frac{\hbar k}{m}\iint d^2\hat{k}d^2\hat{k^\prime} \abs{ f_\Omega(k\hat{k^\prime},k\hat{k})-f_{\Omega^\prime}(k\hat{k^\prime},k\hat{k}) }^2,
\end{equation}
\end{widetext}
where $\frac{\hbar k}{m}$ is the environment particle velocity. Thus, we derive the general expression for the decoherence rate of a quantum rotational system from the elastic scattering model. Equations (\ref{e21}) and (\ref{e22}) are our main results in this section. The expression is general for elastic scattering, since we have not specified any concrete form of the scattering amplitude. Taking into account the spherical symmetry of the environment, we could further rewrite Eq. (\ref{e22}). Denote the scattering amplitude as
\begin{equation}
 f_\Omega(k\hat{k^\prime},k\hat{k})=D^\dag_E(\Omega) f(k\hat{k^\prime},k\hat{k})D_E(\Omega).
\end{equation}
$f(k\hat{k^\prime},k\hat{k})$ is the scattering amplitude for a specific configuration of the system, from which the other scattering amplitude $f_\Omega(k\hat{k^\prime},k\hat{k})$ can be obtained by performing a rotation $D_E(\Omega)$. This greatly simplifies our calculations in the following sections. Several operations yield
\begin{widetext}
\begin{equation}
\label{e25}
\Lambda=\frac{N}{2V}\int dk k^2\mu(k)\frac{\hbar k}{m}\iint d^2\hat{k}d^2\hat{k^\prime} \abs{f(k\hat{k^\prime},k\hat{k})-D^\dag_E(\omega)f(k\hat{k^\prime},k\hat{k})D_E(\omega)}^2,
\end{equation}
\end{widetext}
where we define $D_E(\omega)=D^\dag_E(\Omega^\prime)D_E(\Omega)$, and $\omega$ can be interpreted as the absolute angle distance between the two rotational configurations. Equation (\ref{e25}) shows that the decoherence rate only depends on the absolute angle difference $\omega$ of the configurations, which must result for the case for a spherically symmetric environment. Equation. (\ref{e25}) could greatly simplify our following evaluations since we can always fix one configuration of the system, and fully use its possible symmetry when choosing the coordinates. The other configuration is obtained by just rotating the absolute angle $\omega$ from the fixed configuration. 

\subsection{Comparison with translational decoherence}

In this subsection, we briefly compare the results of the preceding section to the decoherence of the translational degrees of freedom. Recall the case of collisional decoherence for a system's translational degrees of freedom \cite{MS,AS,HS,ASL}, the system density matrix exponentially decays in terms of time 
\begin{equation}
\frac{\partial\rho_S(\vec{x},\vec{x}^\prime,t)}{\partial t}\propto \ -\Lambda\rho_S(\vec{x},\vec{x}^\prime,t).
\end{equation}
The decoherence factor $\Lambda$ is given by
\begin{widetext}
\begin{equation}
\label{e28e}
\Lambda=\frac{N}{2V}\int dk k^2\mu(k)\frac{\hbar k}{m}\iint d^2\hat{k}d^2\hat{k^\prime} \abs{ f_{\vec{x}}(k\hat{k^\prime},k\hat{k})-f_{\vec{x}^\prime}(k\hat{k^\prime},k\hat{k}) }^2,
\end{equation}
\end{widetext}
where $f_{\vec{x}}(k\hat{k},k\hat{k}^\prime)=e^{i\vec{k}\vec{x}}f(k\hat{k},k\hat{k}^\prime)e^{-i\vec{k}^\prime\vec{x}}$. The above formula (\ref{e28e}) takes a similar form as the Eq. (\ref{e22}), where the decoherence rate of the system density matrix depends on the difference of the scattering amplitudes at different values of the pointer variable. The difference of the scattering amplitudes quantifies the distance measure in translational or orientational space. After some operations, one can show that the above decoherence factor is equivalent to 
\begin{widetext}
\begin{equation}
\Lambda=\frac{N}{V}\int dk k^2\mu(k)\frac{\hbar k}{m}\iint d^2\hat{k}d^2\hat{k^\prime} \left(1-e^{ik(\hat{k}-\hat{k}^\prime)(\vec{x}-\vec{x}^\prime)} \right)\abs{f(k\hat{k},k\hat{k}^\prime)}^2,
\end{equation}
\end{widetext}
which is the familiar form for the translational decoherence rate. More details can be found in Ref. \cite{HS,ASL}. In the long wavelength limit, one finds that the translational decay rate is proportional to the position difference square. Similarly, we are expecting the rotational decoherence rate to depend on the angular distance in corresponding orientational space, which is shown in the following sections.

\section{Decoherence due to scattering of thermal photons and mass particles}

In this section, we will explore the theory of rotational decoherence by calculating the decoherence rate for two different sources of decoherence: thermal photons and mass particles.

\subsection{Thermal photon scattering}

We first consider a dielectric ellipsoid immersed in a photon-gas environment. Assuming black-body radiation at temperature $T_E$, the average number of photons with energy $\hbar ck$ is given by the Planck distribution
\begin{equation}
\braket{n(k)}_T=\frac{2}{\exp(\frac{\hbar ck}{k_BT_E})-1},
\end{equation}  
where $c$ is the speed of light. Thus the probability distribution of $k$ with $N$ photons in volume $V$ is 
\begin{equation}
\label{e27}
\mu(k)=\frac{V}{N}\frac{2}{\exp(\frac{\hbar ck}{k_BT_E})-1}.
\end{equation}
To get the decoherence rate, a key task is to evaluate the scattering amplitude difference. For a dielectric object, the cross section is determined by the scattered radiation from the induced dipole. Detailed discussion can be found in Ref. \cite{JDJ}. If we have an incoming field $\vec{E}_{inc}=\vec{\xi}E\exp(-ik\hat{k}\cdot \vec{r})$ with a polarization vector $\vec{\xi}$, the far field approximation gives a scattering amplitude
\begin{equation}
\label{e28}
f(k\hat{k^\prime},k\hat{k})=\frac{k^2}{4\pi\epsilon_0E}\vec{\xi^\prime}\cdot\vec{p},
\end{equation}
where $\vec{\xi^\prime}$ is the polarization of the outgoing radiation, and $\vec{p}$ is the induced dipole moment. The induced dipole moment is given by
\begin{equation}
\label{e29}
\vec{p}=\bar{\bar{\alpha}}_\Omega\cdot \vec{E}_{inc},
\end{equation}
where $\bar{\bar{\alpha}}_\Omega$ is the polarizability of the ellipsoid with configuration $\Omega=(\alpha,\gamma,\beta)$. According to Eq. (\ref{e25}), we can always choose a configuration with the semi-axis of the ellipsoid aligned with the coordinate axis, such that the polarizability is diagonal
\begin{equation}
\label{e30}
\bar{\bar{\alpha}}_{0}=\left(\begin{array}{ccc}\alpha_x & 0 & 0 \\0 & \alpha_y & 0 \\0 & 0 & \alpha_z\end{array}\right),
\end{equation}
where the subscript means the Euler angles are zero for this situation. Then the polarizability with any configuration can be easily derived through the following rotation 
\begin{equation}
\label{e31}
\bar{\bar{\alpha}}_{\Omega^\prime}=R^{\dag}(\Omega^\prime)\bar{\bar{\alpha}}_0 R(\Omega^\prime).
\end{equation}
Now we can calculate the difference of the scattering amplitudes in Eq. (\ref{e25}). Through combining the Eq. (\ref{e28}), (\ref{e29}), (\ref{e30}) and (\ref{e31}), the integral becomes
\begin{widetext}
\begin{equation}
\Lambda=\frac{N}{2V}\int dk k^2\mu(k)\frac{\hbar k}{m}\iint d^2\hat{k}d^2\hat{k^\prime} \frac{k^4}{(4\pi\epsilon_0)^2}\abs{\vec{\xi^\prime}\cdot(\bar{\bar{\alpha}}_0-\bar{\bar{\alpha}}_{\Omega^\prime})\cdot\vec{\xi}}^2,
\end{equation}
\end{widetext}
where $\frac{\hbar k}{m}=c$. In order to evaluate the above integral, we should first average over the polarization direction of the incoming and outgoing field. The procedure is simplified by adopting the following useful identity 
\begin{equation}
\sum_\lambda\xi^{(\lambda)}_i\xi^{(\lambda)}_j=\delta_{ij}-\hat{k}_i\cdot\hat{ k}_j,
\end{equation}
where $\lambda$ is the polarization index and $\{\vec{\xi}^{(1)},\vec{\xi}^{(2)},\hat{k}\}$ form a orthogonal basis set. Including the distribution given by Eq. (\ref{e27}), the integral gives the final result
\begin{equation}
\Lambda=6!\frac{c}{36\epsilon_0^2}\left(\frac{k_BT_E}{\hbar c}\right)^7\zeta(7)*\L,
\end{equation}
where 
\[
\zeta(n)=\frac{1}{(n-1)!}\int^\infty_0 d\chi\frac{\chi^{n-1}}{e^{\chi}-1}
\]
is the Riemann $\zeta$-function and $\zeta(7)\simeq1.00835$, $c$ is the speed of light, and the other parameters are
\begin{equation}
\left\{
\begin{aligned}
\L = & A^2 a_1 +B^2 a_2 +C^2 a_3 + AB a_4 + AC a_5 - BC a_6 ,\\
A  =&\alpha_x-\alpha_y, \\
B  =&\alpha_x-\alpha_z, \\
C  =&\alpha_y-\alpha_z, \\
a_1  = & 3 - 3 \cos(2 \alpha) \cos(2 \gamma) - \cos(2 \alpha) \cos(2 \beta) \cos(2 \gamma)\\
    	   & + 4 \cos\beta \sin(2 \alpha) \sin(2 \gamma),\\
a_2  =&  a_3 =2 -\cos(2 \beta), \\
a_4  =& a_5 =2 \cos(2 \alpha) \sin^2\beta + 2 \cos(2 \gamma) \sin^2\beta,&\\
a_6  =&  2\cos(2\beta).
\end{aligned}
\right.
\end{equation}
In this expression, the polarizabilities $\alpha_x$, $\alpha_y$, and $\alpha_z$ should not be confused with the Euler angle $\alpha$ inside the trigonometric functions. The above equation gives the general expression for the decoherence rate of a dielectric ellipsoid, which depends on the Euler angles and the components of polarizability. This expression gives $\L=0$ when all angles are zero as befitting the requirement that decoherence leaves the diagonal elements of the density matrix unchanged.

If the ellipsoid is cylindrically symmetric, the polarizability components $\alpha_x=\alpha_y$. In this case, only one angle dependence is expected in the decoherence rate, and the result emerges from the general expression. Consider a $z$ axis cylindrical symmetrical ellipsoid, we have $\alpha_x=\alpha_y$. The above result can be reduced to
\begin{equation}
\label{e37}
\Lambda=6!\frac{2c}{9\epsilon_0^2}\left(\frac{k_BT_E}{\hbar c}\right)^7\zeta(7) (\alpha_x - \alpha_z)^2 \sin^2\beta,
\end{equation}
where $\beta$ is the difference in angle between the two orientations. First, we see that the decoherence rate only depends on the angular difference between the two orientations; the decay rate depends on the sine square of the difference in angles, and it will get its maximal when $\beta=\pi/2$. This is reasonable because, as $\beta$ increases, the configuration begins to repeat itself when $\beta$ becomes larger than $\pi/2$. Second, the decoherence rate strongly depends on the thermal temperature. Increasing the temperature will greatly suppress quantum coherence. Also, we find that the temperature dependence for rotational decoherence is two powers lower in $T_E$ than that for center of mass decoherence \cite{MS}. Third, if we totally symmetrize the system by setting $\alpha_x=\alpha_y=\alpha_z$, the orientational decoherence rate will equal to zero because the photon scattering can not distinguish the rotational state of a sphere.

For a specific example, we consider an ellipsoidal nano-diamond, with the size about $100nm$. The nano-diamond shape typically is not elliptical but the following is an estimate and in the long wavelength limit the precise shape isn't important. In the evaluation, we pick $50nm$ and $75nm$ respectively as the short and long half axis of the nano-diamond, which gives ellipticity $e=0.75$. The polarizability satisfies  \cite{BCHD}
\begin{equation}
\alpha_i\sim \epsilon_0 V\frac{\epsilon_d-\epsilon_0}{\epsilon_0+L_i(\epsilon_d-\epsilon_0)}, 
\end{equation}
where $\epsilon_d\sim 6\epsilon_0$ and $\epsilon_0$ are the diamond, vacuum dielectric constant respectively, and $V$ is the volume of the diamond. $L_i$, where the index $(i=x,y,z)$, is determined by the ellipticity of the nano-diamond. For a ellipsoidal diamond with ellipticity $e=0.75$, one can get $L_z\sim0.23$, and $L_x=L_y\sim0.38$. Thus, the decoherence rate is approximated by $\Lambda\sim3.2\times10^{-14}(\frac{T_E}{K})^7\sin^2\beta$ (1/s). 

Refer to the translational degrees of freedom \cite{EH}, the decoherence rate is given by
\begin{equation}
\mathscr{L} =8!\frac{1}{2\pi^3}V^2c(\frac{\epsilon_d-\epsilon_0}{\epsilon_d+2\epsilon_0})^2(\frac{k_BT_E}{\hbar c})^9\zeta(9)\Delta x^2,
\end{equation} 
where $\zeta(9)=1.002$ and $V$ is the volume of the system particle. It is obvious that the ratio of rotational decoherence rate to translational decoherence rate is proportional to $(\frac{\hbar c}{k_B T_E r})^2 \sim \frac{1}{(k_{th} r)^2} \gg 1$, where $k_{th}$ is a thermal photon wave number. In order to roughly compare the rotational and translational decoherence rates, we also pick a nano-diamond with a radius $r\sim50nm$. The spacial separation is chosen $\Delta x=r\sin\beta$, which equals to the "distance" the tip of the ellipsoid move. Thus, the translational decoherence rate is approximated by $\mathscr{L}=1.5\times10^{-23} (\frac{T_E}{K})^9 \sin^2\beta$ (1/s). 

The following table shows the rotational ($1/ \Lambda$) and translational ($1/ \mathscr{L}$) decoherence time scales for several thermal temperatures. First, one can see that the coherence time drops dramatically as the temperature grows from a cold environment to room temperature. Second, for the same thermal temperature, the translational coherence time is much longer than that for rotational degrees of freedom.  The reason is that, when a nano-diamond is moved a distance $x$ having fixed orientation, the scattering hardly changes. However, if we rotate the nano-diamond, the pattern of scattering changes dramatically. Therefore, it is easier to entangle photons with rotations than with translations.

\begin{table}[htdp]
\caption{Estimates of the rotational ($1/\Lambda$) and translational ($1/\mathscr{L}$) decoherence time for different temperatures. The anglular separation $\beta=\pi/20$.}
\begin{center}
\begin{tabular}{|c|c|c|}
\hline
Temperature (K) &  Rotation (sec) & Translation (sec)\\
\hline
3K & $10^{11}$ & $10^{20}$ \\
50K & $10^3$ & $10^9$\\
100K & 10 & $10^6$ \\
200K & $10^{-1}$ & $10^{3} $ \\
300K & $10^{-3}$ & $10^{2}$\\
\hline
\end{tabular}
\end{center}
\label{default}
\end{table}

At last, it is worth mentioning that rotational decoherence is present whenever the different axes have different polarizabilities. A spherical birefringent, dielectric will have different polarizabilities $\alpha_i$ in different directions, leading to a non-zero decoherence rate in Eq. (\ref{e37}). Thus, birefringence can been used to control nanoparticles \cite{AMD,LAM} but will also lead to rotational decoherence from the asymmetrically scattered photons as discussed in this section.

\subsection{Mass particles scattering}

In this subsection, we will consider mass particles as the source of decoherence in an ellipsoidal system. These types of mass particles typically have a very short de Broglie wavelength, for example, the $O_2$ molecule at room temperature has a de Broglie wavelength $\lambda_d\sim10^{-11}m$. The size of a system, such as a nano-diamond or dust particles, is much larger than $\lambda_d$, which trivially indicates that one single scattering could carry away a maximum orientational information. However, as pointed out in Ref. \cite{MS,HS}, it is necessary to determine the lower-bound of the decoherence rate. So we still employ the long wave approximation in the following evaluation. The mass particles are assumed to be in thermal equilibrium which gives the Maxwell-Boltzmann distribution
\begin{equation}
\label{4.36}
\mu(k)=\left(\frac{\hbar^2}{2\pi m k_B T}\right)^{3/2}\exp(-\frac{\hbar^2 k^2}{2mk_BT}).
\end{equation}
Next, we will adopt the Born approximation to evaluate the scattering amplitude. In the Born approximation, the scattering amplitude is given by the following formula
\begin{equation}
\label{e38}
f(\vec{k^\prime},\vec{k})=-\frac{m}{2\pi\hbar^2}\int d^3\vec{r}\exp\big\{-i(\vec{k^\prime}-\vec{k})\vec{r}\big\}*V(\vec{r}),
\end{equation}
in which $V(\vec{r})$ is the potential of the system. In a real situation, the potential can be very complicated. Here, we consider a cylindrical symmetric ellipsoid which is modeled by the following potential
\begin{equation}
V(\vec{r})=D^\dag(\Omega)V_0(\vec{r})D(\Omega),
\end{equation}
where $V_0(\vec{r})=V_0\exp\big\{-a ({x}^2+{y}^2)-b{z}^2\big\}$, with its symmetric axis placed at the $z$ direction. The parameters $a$ and $b$ are positive and unequal. Due to the cylindrical symmetry, two Euler angles are enough to specify the orientation. The symmetric axis of the potential $V(\vec{r})$ is in any direction determined by the Euler angles. We first calculate the scattering amplitude Eq. (\ref{e38}) with the above potential ${V}(\vec{r})$. A convenient way to do the integral is in Cartesian coordinates. We first perform a coordinate rotation $O(x,y,z)\rightarrow \widetilde{O}(\tilde{x},\tilde{y},\tilde{z})$ to get $V(\vec{r})\rightarrow \widetilde{V}(\vec{r})$, such that the symmetric axis of the potential $\widetilde{V}(\vec{r})$ is aligned with the $\tilde{z}$ axis, then calculate the integral in the $\widetilde{O}$ coordinate. At last, the final scattering amplitude is obtained by rotating the integral result back to the original coordinate $O$. Finally, the scattering amplitude is given by
\begin{equation}
\label{4.40}
f_{\alpha,\beta}(\vec{k^\prime},\vec{k})=\frac{mV_0}{2\pi\hbar^2}\frac{\pi}{a}\sqrt{\frac{\pi}{b}}
\exp\bigg\{
-\frac{\widetilde{\Delta k_x}^2}{4a}-\frac{\widetilde{\Delta k_y}^2}{4a}-\frac{\widetilde{\Delta k_z}^2}{4b}
\bigg\},
\end{equation}
where the vectors $\overrightarrow{\widetilde{\Delta k}}=(\widetilde{\Delta k_x},\widetilde{\Delta k_y},\widetilde{\Delta k_z})$ and $\overrightarrow{\Delta k}=(\Delta k_x,\Delta k_y,\Delta k_z)$ satisfy $\overrightarrow{\widetilde{\Delta k}}=R^{-1}_y(\beta)R^{-1}_z(\alpha)\overrightarrow{\Delta k}$, and $\Delta k_i=k_i^\prime-k_i$, $\widetilde{\Delta k_i}=\widetilde{k_i^\prime}-\widetilde{k_i},$ ($i=x,y,z$). In the following evaluation, we can Taylor expand Eq. (\ref{4.40}) and keep the first order term in the long wavelength limit. Substitute Eq. (\ref{4.36}) and Eq. (\ref{4.40}) into Eq. (\ref{e25}), we get
\begin{equation}
\Lambda=\frac{32\pi}{15\hbar^{8}}N/V\sqrt{2\pi m^7(k_b T)^5}\frac{(a-b)^2V_0^2}{a^4b^3}\sin^2\beta.
\end{equation}
First, we see that the decoherence rate only depends on the polar angle, which specifies the angle difference for the current situation. Second, the parameter $\frac{(a-b)^2V_0^2}{a^4b^3}$ is determined by the size and geometry of the system. When we set $a=b$, the system becomes spherically symmetric, which reduces the decoherence rate to zero. Moreover, the rate has a dependence on the two and a half power of the temperature, which is one power higher than that for the case of center of mass decoherence \cite{MS}. At last, the rate is also proportional to environment particle density $N/V$, which is quite reasonable because higher density increases the scattering rate.

\section{Conclusion}

Decoherence, since the early 80s, has been used to study a vast array of phenomena ranging from microscopic to cosmological scales \cite{HH}. Instead of the Schr$\ddot{\text{o}}$dinger equation, the evolution of an open system is described by a quantum master equation. For different situations, the master equation can be simplified by employing different models, such as quantum brownian motion, spin-boson interactions etc \cite{LCSD,BHPF}. The decoherence, as well as dissipation, appears naturally in the equation. For decoherence, much effort has been devoted to obtain the pointer states of a given master equation, which is important in establishing the quantum to classical transition. Meanwhile, using a scattering model \cite{HS,ASL}, the decoherence effect is widely studied for center of mass motions, where the environmentally distinguished states become exponentially localized wave packets \cite{MS, HS}. 

In this paper, based on a scattering model, we show that the same decoherence effect holds for a quantum system with rotational degrees of freedom. As we have shown in this paper, a quantum system with rotational degrees of freedom suffers from decoherence when interacting with an external environment. The environment of photon gas or mass particles is able to exponentially localize the rotational state, with the decoherence rate proportional to the difference of rotational configurations. Interestingly, the decay rate has a temperature dependence that is different from that for translational decoherence. The rotational decoherence due to photons seems to be faster than translational decoherence by a factor of $1/(k_{th} r)^2 \gg 1$, where $k_{th}$ is a thermal photon wave number and $r$ is the particle size.

The study of rotational decoherence is not only fundamentally meaningful, but also instrumental for the growing interests in accurate quantum control over a system's internal motion. As more experimental evidences for mesoscopic quantum phenomena are found, to consider the decoherence in all degrees of freedom becomes critical. In Ref. \cite{TL}, the decoherence of center of mass motion induced the interaction with a system's own internal degree of freedom is even suggested. For any accurate quantum control or the manufacture of quantum devices, it is extremely important to identify the decoherence mechanism and the corresponding decoherence time. In conclusion, the study of rotational decoherence, together with decoherence with other degrees of freedom, will surely contribute as a useful guidance to future mesoscopic-scale experiments and applications.

The author thanks Prof. K. Hornberger for providing Ref. \cite{TF} and Prof. Tongcang Li for helpful discussions. This work was supported by the National Science Foundation under Grant No.1404419-PHY.

\bibliographystyle{ieeetr}
\bibliography{all.bib}

\end{document}